\documentclass[12pt]{article}
\usepackage{graphicx}
\usepackage{color}
\usepackage{cite}
\usepackage{amsmath}
\usepackage{here}

\def\css{\hbox{$c\kern0.03em(\kern-0.1em ss\kern-0.1em)$}}
\def\bss{\hbox{$b\kern0.03em(\kern-0.1em ss\kern-0.1em)$}}
\def\qss{\hbox{$q\kern0.03em(\kern-0.1em ss\kern-0.1em)$}}

\def \beq{\begin{equation}}
\def \eeq{\end{equation}}
\def\eqref#1{(\ref{#1})}
\def\bea{\begin{eqnarray}}
\def\eea{\end{eqnarray}}

\def\jpsi{J\kern-0.15em/\kern-0.15em\psi\kern0.15em}

\def\Sd{S_{\kern-0.1em ss}}
\def\SSd{{\bold S}_{\kern-0.1em ss}}
\def\Sdi{S^i_{\kern-0.1em ss}}

\def\URLtilde{\lower0.2em\hbox{$\tilde{\phantom{a}}$}}
\def\mycomm#1{\hfill\break\strut\kern-3em{\color{red}\tt ====> #1
\color{black}}\hfill\break}

\newcount\timecount
\newcount\hours \newcount\minutes  \newcount\temp \newcount\pmhours
\hours = \time
\divide\hours by 60
\temp = \hours
\multiply\temp by 60
\minutes = \time
\advance\minutes by -\temp
\def\hour{\the\hours}
\def\minute{\ifnum\minutes<10 0\the\minutes
\else\the\minutes\fi}
\def\clock{
\ifnum\hours=0 12:\minute\ AM
\else\ifnum\hours<12 \hour:\minute\ AM
\else\ifnum\hours=12 12:\minute\ PM
\else\ifnum\hours>12
\pmhours=\hours
\advance\pmhours by -12
\the\pmhours:\minute\ PM
\fi
\fi
\fi
\fi
}

\def\monthname{\relax\ifcase\month 0/\or January\or February\or
March\or April\or May\or June\or July\or August\or September\or
October\or November\or December\else\number\month/\fi}
\def\today{\monthname~\number\day, \number\year}

\def\bold#1{\boldsymbol{#1}}

\def\draft{\color{red}
$\bold{\strut\kern-3em
\hbox{\tt \Large DRAFT, NOT TO BE DISTRIBUTED:  \clock, \today.}
}$\par\noindent\color{black}}
\begin{document}
\setcounter{footnote}{1}
\rightline{EFI 23-X}
%\rightline{TAUP YYYY/23}
\rightline{arXiv:2305.00354}
\vskip -0.3cm
\centerline{\bf SINGLE-PHOTON DECAYS}
\medskip

\centerline{\bf IN SYSTEMS WITH AT LEAST ONE HEAVY QUARK}
\unboldmath
\bigskip

 \centerline{Marek Karliner$^a$\footnote{{\tt marek@tauex.tau.ac.il}},
and Jonathan L. Rosner$^b$\footnote{{\tt rosner@hep.uchicago.edu}}}
\medskip

\centerline{$^a$ {\it School of Physics and Astronomy}}
\centerline{\it Raymond and Beverly Sackler Faculty of Exact Sciences}
\centerline{\it Tel Aviv University, Tel Aviv 69978, Israel}
\medskip

\centerline{$^b$ {\it Enrico Fermi Institute and Department of Physics}}
\centerline{\it University of Chicago, 5640 S. Ellis Avenue, Chicago, IL
60637, USA}
\strut

\begin{center}
ABSTRACT
\end{center}
\begin{quote}
Hadrons containing at least one heavy quark (charm or bottom) frequently
have small enough natural widths that decay modes involving a single photon
have detectable branching fractions.  Photons of typical energy greater than
100 MeV have been directly detected, while those of lower energy have only
been inferred.  Here we discuss prospects for observing direct sub-100 MeV
photons in specific radiative decays of charmed and bottom vector mesons,
as well as a spin-excited heavy baryon.
\end{quote}
\smallskip

\section{{\bf Introduction} \label{intro}}

The spectroscopy of mesons and baryons containing at least one heavy quark
($c$ or $b$) has blossomed in recent years.  In part this is because
many newly discovered states have had small natural widths, tens of MeV or
less, allowing single-photon emission to compete favorably with hadronic decay.
The reduction of partial widths is further enhanced if the heavy hadron
contains {\it no} light quarks ($u$ or $d$), because coupling to pions then
is suppressed.

Table \ref{tab:vpg} lists some decays of interest here.  The single photons
of energy typically greater than 100 MeV have been directly detected, while
those of energy less than 100 MeV (marked with an asterisk) mostly remain to be
observed.  Many are produced in decays of vector mesons $V$ to pseudoscalar
mesons $P$ and single photons.  Their low energies are due to the small
hyperfine splittings between the initial $V$ and final $P$. Table \ref{tab:bbg}
gives the corresponding data for the charmed-strange baryons $\Xi_c^{+,0}$.
All data are quoted from PDGLive \cite{PDG} except where indicated.
The rare cases in which a direct photon of less than 100 MeV {\it is} observed
(e.g., in $\Upsilon(3S) \to \chi_{b1,2}(2P) \gamma$) are due to low backgrounds
and good final-state reconstruction.

% This is Table I
\begin{table}
\caption{Some $X \to Y \gamma$ transitions for states with at least one
heavy quark. 
\hfill\break
Asterisks denote decays not yet observed directly.\label{tab:vpg}}
\begin{center}
\begin{tabular}{c c c c} \hline \hline
$X$ & $Y$ & $E_\gamma$ & BR \\
    &     &    (MeV)   & (\%) \\ \hline
$D^{*0}$ & $D^0$ & 137 & $35.3 \pm 0.9$ (a) \\
$D^{*+}$ & $D^+$ & 136 & $1.6 \pm 0.4$ (a) \\
$D_s^{*+}$ & $D_s^+$ & 139 & $93.5 \pm 0.7$ \\
$B^*$ & $B$ & 45$^*$ & 100 \\
$B_s^{*0}$  & $B_s^0$ & 48$^*$ & 100 \\
$B_c^{*+}$ & $B_c^+$ & 54--84$^*$ (b) & 100 \\
$J/\psi$  & $\eta_c$  & 111 & $1.7\pm0.4$ \\
$\Upsilon(1S)$ & $\eta_b$ & $61 \pm 2^*$ & (c) \\
$\Upsilon(2S)$ & $\eta_b$ & 605 & $(5.5^{+1.1}_{-0.9})\times 10^{-4}$ \\
$\Upsilon(2S)$ & $\chi_{b0}$ & 162 & $3.8 \pm 0.4$ \\
$\Upsilon(2S)$ & $\chi_{b1}$ & 130 & $6.9 \pm 0.4$ \\
$\Upsilon(2S)$ & $\chi_{b2}$ & 110 & $7.15\pm0.35$ \\
$\Upsilon(3S)$ & $\chi_{b0}(2P)$ & 122 & $5.9 \pm 0.6$ \\
$\Upsilon(3S)$ & $\chi_{b1}(2P)$ & 99 & $12.6 \pm 1.2$ \\
$\Upsilon(3S)$ & $\chi_{b2}(2P)$ & 86 & $13.1 \pm 1.6$ \\
\hline \hline
\end{tabular}
\end{center}
{\footnotesize
\leftline{\strut\kern6em
(a) In $D^{*0} \to D^0 \gamma$ the contributions of the charmed and
light quarks add}
\leftline{\strut\kern7.6em 
constructively, while they nearly cancel one another in $D^{*+}
\to D^+ \gamma$.}
\leftline{\strut\kern6em
\vrule width 0pt height 2.8ex
(b) To be determined in this work; our present estimate of range.}	
\leftline{\strut\kern6em
\vrule width 0pt height 2.8ex
(c) The branching fraction to three gluons is about 82\%, while that to}
\leftline{\strut\kern7.65em 
each lepton pair is about 2.5\% and to $\gamma gg$ is about 
2.2\%. Thus the}
\leftline{\strut\kern7.65em 
branching fraction for $\Upsilon(1S) \to \eta_b \gamma$,
though not quoted, is expected}
\leftline{\strut\kern7.65em 
to be considerably less than 1.}
} %end of \footnotesize
\end{table}
%
%This is Table II
\begin{table}
\caption{Single-photon emission by baryons.  States with at least one heavy
quark.  \label{tab:bbg}}
\begin{center}
\begin{tabular}{c c c c} \hline \hline
Initial & Final  & $E\gamma$ & BR \\
 baryon & baryon &   (MeV)   & (\%) \\ \hline
$\Xi_c^{'+}$ & $\Xi_c^+$ & 108 & 100 \\
$\Xi_c^{'0}$ & $\Xi_c^0$ & 106 & 100 \\
\hline \hline
\end{tabular}
\end{center}
\end{table}
Some predicted M1 transition rates are summarized in Table \ref{tab:m1rat}.
% This is Table III
\begin{table}
\caption{M1 transition rates and branching ratios for $M^* \to M \gamma$
decays:  predicted [Eq.\ (1) with $|I|^2 = 1/2$] and (where available)
observed \cite{PDG}.  Quark masses are taken to be $m_u = m_d = 308.5$ MeV,
$m_s = 482.2$ MeV, $m_c = 1665$ MeV, $m_b = 5041$ MeV.
\label{tab:m1rat}}
\begin{center}
\begin{tabular}{c c c c c c} \hline \hline
  Quarks     &    &    & Rate $\Gamma$ & \multicolumn{2}{c}{Branching ratio} \\
$q_1\bar q_2$ & $M^*$ & $M$ & Pred (keV) & Pred & Obs \\ \hline
$c \bar u$ & 2006.8  & 1864.8 & 20.54 & (a) & (35.3$\pm$0.9)\% \\
$c \bar d$ & 2010.3  & 1869.7 & 1.40 & (a) & (1.6$\pm$0.4)\% \\
$c \bar s$ & 2112.2  & 1968.3 & 0.272 & & (93.5$\pm$0.7)\% \\
$c \bar c$ & 3096.9  & 2983.9 & 1.077 & 1.16\% & (1.7$\pm$0.4)\% \\
$b \bar d$ & 5324.7  & 5279.7 & 0.144 & -- & 100\% \\
$b \bar u$ & 5324.7  & 5279.3 & 0.491 & -- & 100\% \\
$b \bar s$ & 5415.4  & 5366.9 & 0.0787 & -- & 100\% \\	
$b \bar c$ & 6343.5  & 6274.5 & 0.0443 & -- & 100\% \\
$b \bar b$ & 9460.3  & 9398.7 & 0.00503 & $9.3 \times 10^{-5}$ & -- \\ 
\hline \hline
\end{tabular}
\end{center}
\leftline{(a) See Eqs.\ (\ref{eqn:dsumt},\ref{eqn:dsumx}) for prediction of
hadronic and radiative $D^*$ widths.}
\end{table}

We shall describe magnetic dipole (M1) radiative transitions by sums of quark
spin-flip amplitudes.  The decays of vector mesons $M^*$ to pseudoscalar mesons
$M$ composed of quarks $q_1 \bar q_2$ proceed at a rate \cite{Eichten:1978tg}
\beq
\Gamma(M^* \to M \gamma) = \frac{p_\gamma^3}{3\pi} [\mu_1 + \mu_2]^2 |I|^2~,
\eeq
where $p_\gamma$ is the photon momentum in the c.m.s., $\mu_1$ and $\mu_2$ are
the magnetic moments of the quarks, and $\mu_i = eQ_i/(2m_i)$, with $eQ_i$
denoting the charge of quark $i$. The amplitude $I=\langle f|i \rangle$ denotes
the overlap between initial and final wave functions.  For a wide range of
cases, involving both light and heavy quarks, $|I|^2 \simeq 1/2$.
(For a summary of early work on radiative decays of vector mesons to
pseudoscalar mesons see Ref.\ \cite{Casalbuoni:1996pg}.)

In this paper we discuss status and prospects for observing direct photons in
$D^* \to D \gamma$ and $D_s^* \to D_s \gamma$ in Sec.\ \ref{sec:dg},
$B^* \to B \gamma$ and $B_s^* \to B_s \gamma$ in Sec.\ \ref{sec:bg},
$\Upsilon(1S) \to \eta_b \gamma$ in Sec.\ \ref{sec:bot}, $\Xi_c' \to \Xi_c
\gamma$ in Sec.\ \ref{sec:xic} and $B_c^* \to B_c \gamma$ in Sec.\
\ref{sec:bc}, concluding in Sec.\ \ref{sec:con}.  An Appendix
compares amplitudes to M1 transitions.

\section{$D^* \to D \gamma$ and $D_s^* \to D_s \gamma$ \label{sec:dg}}

\subsection{Nonstrange $D$ mesons}

The spacing between the pseudoscalar and vector nonstrange $D$ mesons is just
large enough to accommodate single pion emission, leading to hadronic decays
which can compete with single photon emission.  The pattern of $D$ decay widths
was analyzed in Ref.\ \cite{Rosner:2013sha}, with successful descriptions of
hadronic \cite{Amundson:1992yp} and radiative (M1) transitions.
Combining hadronic and radiative transitions, it was predicted that
\beq \label{eqn:dsumt}                   
\Gamma_{\rm tot}(D^{*+}) = (79.0 + 2.76|I|^2)~{\rm keV}~~~,~~~
\Gamma_{\rm tot}(D^{*0}) = (34.7 + 40.8|I|^2)~{\rm keV}~,
\eeq
to be compared with experimental values \cite{PDG}
\beq \label{eqn:dsumx}
\Gamma_{\rm tot}^{\rm}(D^{*+}) = (83.4 \pm 1.8)~{\rm keV}~~~,~~~
\Gamma_{\rm tot}^{\rm}(D^{*0} < 2.1~{\rm MeV}~.
\eeq
Note in particular the relatively large value of the partial width for
$D^{*0} \to D^0 \gamma$ quoted in Table \ref{tab:m1rat}.  Using the expression
in Eq.\ (2) for $\Gamma_{\rm tot}(D^{*0}$, and the observed branching fraction
${\cal B}(D^{*0} \to D^0 \gamma) = 35.3\%$, one finds $|I|^2 = 0.576$ and
$\Gamma_{\rm tot}(D^{*0}) = 58.2$ keV.

\subsection{Strange $D$ mesons}

The hadronic decay of $D_s^{*+}$ to $\pi \pi D_s^+$ is forbidden kinematically
and to $\pi^0 D_s^+$ is forbidden by isospin.  The dominant decay is then
$D_s^* \to D_s \gamma$, accounting for a branching fraction of $(93.5 \pm
0.07)\%$ and hence a total width of $(291 \pm 8)$ eV.  A determination of this
quantity in lattice QCD \cite{Donald:2013sra} finds a much smaller value,
$\Gamma_{\rm tot}(D_s^*) = (70 \pm 28)$ eV, which has been combined with an
experimental study of the purely leptonic decay $D_s^{*+} \to e^+\nu_e$ to
obtain a first value of the $D_s^{*+}$ decay constant \cite{BESIII:2023zjq}. 
It would be interesting to see if a similar discrepancy between predicted and
observed total widths holds for the $D^{*0}$ decays.

\section{$B^* \to B \gamma$ and $B_s^* \to B_s \gamma$ \label{sec:bg}}

The two charge states in $B^* \to B \gamma$ are not separated from
one another \cite{PDG}, so one is missing valuable information on the
masses of $B^{*0}$ and $B^{*+}$.  This in turn would add data to systematics
of isospin splittings \cite{Karliner:2019lau}.

The detection of a photon of 45 MeV in $B^* \to B \gamma$ or 48 MeV in
$B_s^* \to B_s \gamma$ poses a significant challenge.  As in all processes
discussed here and below, reconstruction of the final state is essential.
If there are enough events, one may gain background suppression by asking
for a Dalitz decay in which the photon materializes as an $e^+ e^-$ pair.
Furthermore, a boosted decay geometry such as is available in LHCb may
compensate for the very low photon energy in the center-of-mass (c.m.).
On the other hand, fully reconstructed final states in $e^+e^-$ reactions,
such as provided by Belle II at a c.m. energy of the $\Upsilon(4S)$, may
have backgrounds low enough to permit detection of the $\sim 45$ MeV
photons.

\section{$\Upsilon(1S) \to \eta_b \gamma$ \label{sec:bot}}

The $\eta_b(1S)$ was not discovered through the decay $\Upsilon(1S) \to
\eta_b \gamma$, which requires detection of a 61 MeV photon.  It was seen with
highest statistics by Belle and BaBar in the decay $\Upsilon(2S) \to \eta_b
\gamma$ \cite{PDG} (see $\Gamma_{75}$), in which $E_\gamma \simeq 605$ MeV.  
Adding to the
difficulty is the absence of a large, easily reconstructed $\eta_b$ branching
fraction.  The best route to detection of the 61 MeV photon would be to study
the exclusive final states $X$ in the {\it observed} $\eta_b$ decays
and then select them in $\Upsilon(1S) \to X \gamma$. 

\section{$\Xi_c' \to \Xi_c \gamma$ \label{sec:xic}}

The photon in either charge state of $\Xi_c' \to \Xi_c \gamma$ has an energy
of $106$--$108$ MeV.  These decays are marked as ``seen'' in the Particle
Listings \cite{PDG}, based primarily on 2016 Belle data \cite{Belle:2016lhy}. 
Their analysis of the decays $\Xi_c' \to \Xi_c \gamma$ is a textbook example
of the care that must be taken in detecting a low-energy photon.  The
strength of the observed signals suggests sensitivity to photon energies
well below 100 MeV.

\section{$B_c^* \to B_c \gamma$ \label{sec:bc}}

The study of the $b \bar c$ system is hampered by ignorance of the splitting
$\Delta_{1S} \equiv M[B_c^*(^3S_1)]-M[B_c(^1S_0)]$.  Some early estimates
include ones quoted in Table \ref{tab:mbc}, while some recent ones 
include 54 MeV \cite {Eichten:2019gig} and 68--84 MeV \cite{Karliner:2014gca}.
We thus take $\Delta_{1S}$ = 54 MeV to 84 MeV, or $69 \pm 15$ MeV.
Splitting between the 1S and 2S states has been studied by ATLAS
\cite{ATLAS:2014lga}, CMS \cite{CMS:2019uhm}, and LHCb \cite{LHCb:2019bem}.
Taking $\Delta_{1S} = 69 \pm 15$ MeV, 
we estimate $M[B^*_c(1S)]$ in the last line of Table \ref{tab:mbc}. 
A feature which distinguishes the CMS result from the others is the
observation of two narrow peaks in the $M(B_c^+\pi^+\pi^-)-M(B_c^+)$ spectrum,
suggesting good final-state reconstruction.  These peaks are separated by
$29.1\pm 1.5 \pm 0.7$ MeV, with the lower most likely corresponding to the
$^3S_1$ states and the higher most likely corresponding to the $^1S_0$ states.%
\footnote{This assumption is motivated by theoretical quark model calculations.}
One can regard the 29.1 MeV as representing the difference between the
hyperfine splittings of the 1S and 2S $B_c$ doublets.  One uses this quantity
in estimating $M[B^*_c(2S)]$ in the last line of Table \ref{tab:mbc}.
Detection of the photon in $B^*_c(1S) \to B_c(1S) \gamma$ thus would be a
useful adjunct to the spectroscopy of these states.

% This is Table IV
\begin{table}
\caption{Some early and current predictions of 1S and 2S $b \bar c$ masses
(MeV) compared with present values ($M(B_c)$) or estimates ($M(B_c^*)$).
\label{tab:mbc}}
\begin{center}
\begin{tabular}{c c c c c} \hline \hline
Reference & $M[B_c(1S)]$ & $M[B^*_c(1S)]$ & $M[B_c(2S)]$ & $M[B^*_c(2S)]$ \\
\hline
Eichten94 \cite{Eichten94} & 6264 & 6337 & 6856 & 6899 \\
Gershtein95 \cite{Gershtein95} & 6253 & 6317 & 6867 & 6902 \\
Fulcher99 \cite{Fulcher99} & 6286 & 6341 & 6882 & 6914 \\
Ebert03 \cite{Ebert03} & 6270 & 6332 & 6835 & 6881 \\
Eichten19 \cite{Eichten:2019gig} & 6275 & 6329 & 6867 & 6898 \\
Present & $6274.47\pm0.32$ & $6343\pm15$ & $6872.1\pm1.0$ & $6912\pm15$ \\
\hline \hline
\end{tabular}
\end{center}
\end{table}

Although the CMS collaboration reports a very small (1\%) reconstruction
efficiency for the soft photon in $B_c^* \to B_c \gamma$, it might be worth
while exploring ways to increase this efficiency.  Possibilities include (i)
selecting a sample of the most highly boosted $B_c$ events, (ii) replacing
the photon with an $e^+ e^-$ pair (Dalitz decays), and (iii) looking for
photons which convert in the material of the detector.

\section{Conclusions \label{sec:con}}
The ability to reconstruct photons of energies less than 100 MeV would
greatly enhance the versatility of detectors studying the spectroscopy of
states with one or more heavy ($c$ or $b$) quarks.  We have discussed ways of
improving their sensitivity, in such decays as $B^* \to B \gamma$, $B_s^*  \to
B_s \gamma$, $\Upsilon(1S) \to \eta_b \gamma$, and $B_c^* \to B_c \gamma$.
These decays are also of interest because they can shed light on the relative
strengths of weak decays.  We thank Tim Gershon for emphasizing this
point, and alerting us to the suggestion of Ref.\ \cite{Grinstein:2015aua}.

\section*{Appendix}

\begin{figure}[H]
\begin{center}
 \includegraphics [width = 0.90\textwidth]{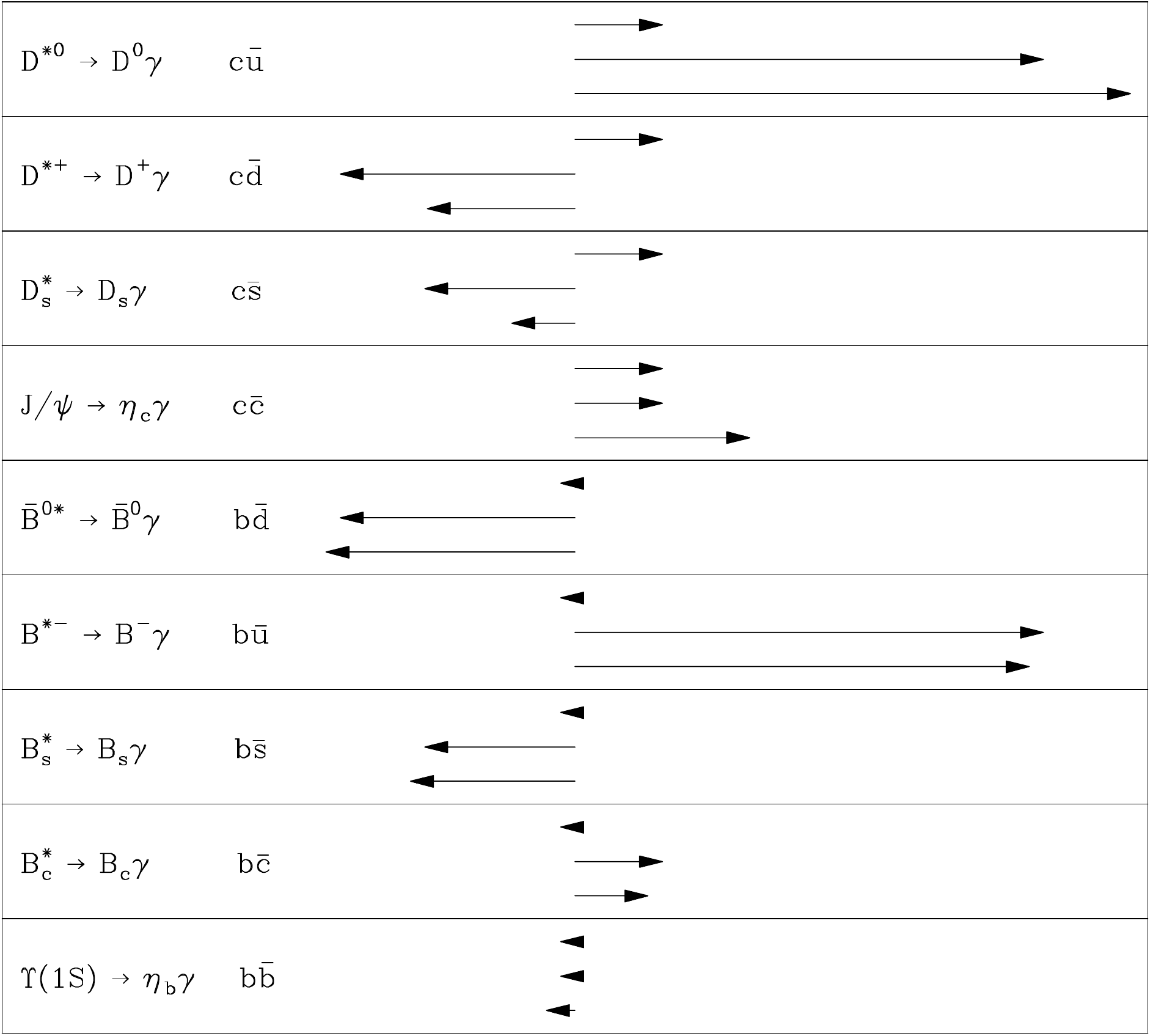}
\end{center}
\caption{%
Graphical depiction of Table \ref{tab:amps}.
Relative amplitudes contributing to magnetic dipole (M1) transitions
of $q_1 \bar q_2$ states from $J^P = 1^-$ to $0^-$.  
In each box the first
arrow denotes the contribution of $q_1$, the second denotes the contribution
of $\bar q_2$, and the third denotes their sum.
\label{fig:amps}}
\end{figure}

The individual contributions of magnetic moments of quarks $q_i$ or antiquarks
$\bar q_i$ are terms $eQ_i/ (2 m_i)$ proportional to magnetic moments which
interfere with one another constructively or destructively in Eq.\ (1).  They
are compared in Table \ref{tab:amps} and Figure \ref{fig:amps}.  
\vfill\eject
%

% This is Table V
\begin{table}
\caption{Amplitudes $\mu = eQ/(2m_q)$ contributing to M1 transitions involving
single-photon decays.  The arrows in each box give the contributions of
quark $q_1$, antiquark $\bar q_2$, and their sum (all in units of $10^{-6}$
(MeV)$^{-1}$).}
\label{tab:amps}
\begin{center}
\begin{tabular}{c c c c c} \hline \hline
Decay & Quarks & $\mu(q_1)$ & $\mu(\bar q_2)$ & $\mu(q_1)+\mu(\bar q_2)$ \\
\hline
$D^{*0} \to D^0 \gamma$ & $c \bar u$ & \phantom{---}61 & 
\phantom{---}327 & \phantom{---}388 \\
$D^{*+} \to D^+ \gamma$ & $c \bar d$ & \phantom{---}61 & $-164$ & $-103$ \\
%$D^*_s \to D_s \gamma $ & $c \bar s$ & \phantom{---}61 & $-105$ & $-43.7$ \\
 $D^*_s \to D_s \gamma $ & $c \bar s$ & \phantom{---}61 & $-105$ & 
\phantom{--}$-44$ \\
$J/\psi \to \eta_c \gamma$ & $c \bar c$ & \phantom{---}61 & 
\phantom{----}61 & \phantom{---}122 \\
$\bar B^{*0} \to \bar B^0 \gamma$ & $b \bar d$ & $-10$ & $-164$ & $\,-174$ \\
$B^{*-} \to B^- \gamma$ & $b \bar u$ & $-10$ & 
\phantom{---}327 & \phantom{---}317 \\
$B^*_s \to B_s \gamma$ & $b \bar s$ & $-10$ & $-105$ & $\,-115$ \\
$B^*_c \to B_c \gamma$ & $b \bar c$ & $-10$ & \phantom{----}61 & 
\phantom{----}51 \\
$\Upsilon(1S) \to \eta_b \gamma$ & $b \bar b$ & $-10$ & \phantom{-}$-10$ 
& \phantom{--}$-20$ \\
\hline \hline
\end{tabular}
\end{center}
\end{table}
%
% This is Figure 1
%
\strut\vskip-2.6cm\strut

\noindent
Several properties are notable.
\hfill\break

(1) The largest amplitude is for $D^{*0} \to D^0 \gamma$, benefitting from a
large radiative contribution from the $u$ quark magnetic moment and 
constructive interference with a $c$ quark moment.  The radiative width
of $D^{*0} \to D^0 \gamma$ then becomes comparable to the hadronic width of
$D^{*0} \to D^0 \pi^0$ \cite{Rosner:2013sha}.
\hfill\break

(2) The $d$ quark moment's contribution to $(D^{*+} \to D^+ \gamma,~\bar B^{*0}
\to \bar B^{*0} \gamma)$ 
is -1/2 times the $u$ quark moment's contribution to
($D^{*0} \to D^0 \gamma,~B^{*-} \to B^- \gamma$), simply reflecting the charge
of the light quark.
\hfill\break

(3) In $D^{*+} \to D^+ \gamma$ the $c$ and $d$ quark moments interfere
destructively.  The combined effect of (2) and (3) is to give a much smaller
radiative width for $D^{*+} \to D^+ \gamma$ than for $D^{*0} \to D^0 \gamma$.
\hfill\break

(4) In comparing $D_s^* \to D_s \gamma$ with $D^{*+} \to D^+ \gamma$, the
percent of U-spin violation in the $s$ vs.\ $d$ contributions is amplified by
destructive interference with the $c$ contribution.  Thus the radiative width
for $D_s^* \to D_s \gamma$ is quite sensitive to the exact pattern of this
interference.
\hfill\break

(5) For example, the measurement of the branching fraction 
${\cal B}(D_s^* \to e^+ \nu_e) = (2.1^{+1.2}_{-0.9}\pm0.2)\% $ 
\cite{BESIII:2023zjq} combined with the
Particle Data Group \cite{PDG} average ${\cal B}(D_s^*\to D_s \gamma) 
= (93.5 \pm 0.7)\%$ and
Eq.\ (1) in the BESIII paper for the partial width $\Gamma(D_s^*
\to e^+ \nu_e)$ gives 
\beq
    \Gamma(D_s^*\to D_s \gamma) = 
\Gamma(D_s^* \to e^+ \nu_e)/{\cal B}(D_s^* \to e^+ \nu_e)\times
{\cal B}(D_s^*\to D_s \gamma) \,=\,
(114^{+65}_{-50})~{\rm eV}~,
\eeq
to be compared with a lattice QCD \cite{Donald:2013sra} estimate 
\beq
\Gamma(D_s^* \to D_s \gamma) = 66 \pm 26~{\rm eV},
\eeq
even smaller than the value of
\beq
   \Gamma(D_s^* \to D_s \gamma) = 272~{\rm eV} |I|^2/(1/2)
\eeq
obtained in the present work (cf.\ 3rd row of Table \ref{tab:m1rat}).
\hfill\break

(6) The magnetic dipole transition amplitude for $B^{*-} \to B^- \gamma$ is
appreciable, not much affected by a small destructive interference 
between the large $u$ quark and small $b$ quark contributions.
\hfill\break

(7) Some of the M1 transition rates we have been discussing, while being
the main
or exclusive decay modes of the vector meson of interest, are sufficiently
low that weak decays can appear with detectable branching ratios.  This has
already been seen by BESIII \cite{BESIII:2023zjq} in the case of $D_s^* \to 
D_s \gamma$ vs.\ $D_s^* \to e^+ \nu_e$, and may also be promising for
$B^{*-}$, $B_s^*$, and $B_c^*$ decays (see Table \ref{tab:m1rat}).

\section*{Acknowledgments}
We thank Christine Davies and Tim Gershon for helpful comments.
The work of M.K. was supported in part by NSFC-ISF Grant No.\ 3423/19.

\end{document}